\documentclass[iop]{emulateapj}

\def\ltsima{$\; \buildrel < \over \sim \;$}
\def\simlt{\lower.5ex\hbox{\ltsima}}
\def\gtsima{$\; \buildrel > \over \sim \;$}
\def\simgt{\lower.5ex\hbox{\gtsima}}
\def\kpc{{\rm\,kpc}}

\def\pc{{\rm\,pc}}

\def\deg{^\circ}

\def\s{\ifmmode \widetilde \else \~\fi}
\def\={\overline}

\def\spose#1{\hbox to 0pt{#1\hss}}

\def\lta{\mathrel{\spose{\lower 3pt\hbox{$\mathchar"218$}}
     \raise 2.0pt\hbox{$\mathchar"13C$}}}
\def\gta{\mathrel{\spose{\lower 3pt\hbox{$\mathchar"218$}}
     \raise 2.0pt\hbox{$\mathchar"13E$}}}
\def\Dt{\spose{\raise 1.5ex\hbox{\hskip3pt$\mathchar"201$}}}    
\def\dt{\spose{\raise 1.0ex\hbox{\hskip2pt$\mathchar"201$}}}    

\def\dotsfill{\leaders\hbox to 1em{\hss.\hss}\hfill}
\def\sun{\odot}

\def\Gyr{{\rm\,Gyr}}
\def\FeH{{\rm[Fe/H]}}

\shorttitle{A new distant distant Milky Way globular cluster}
\shortauthors{B. P. M. Laevens et al.}

\begin{document}


\title{A new distant Milky Way globular cluster in the Pan-STARRS1 $3\pi$ survey}


\author{Benjamin P. M. Laevens$^{1,2}$, Nicolas F. Martin$^{1,2}$, Branimir Sesar$^2$, Edouard J. Bernard$^3$, Hans-Walter Rix$^2$, Colin T. Slater$^4$, Eric F. Bell$^4$, Annette M. N. Ferguson$^3$, Edward F. Schlafly$^2$, William S. Burgett$^5$, Kenneth C. Chambers$^5$, Larry Denneau$^5$, Peter W. Draper$^6$, Nicholas Kaiser$^5$,  Rolf-Peter Kudritzki$^5$, Eugene A. Magnier$^5$, Nigel Metcalfe$^6$, Jeffrey S. Morgan$^5$, Paul A. Price$^7$, William E. Sweeney$^5$, John L. Tonry$^5$,  Richard J. Wainscoat$^5$, Christopher Waters$^5$}

\email{benjamin.laevens@astro.unistra.fr}

\altaffiltext{1}{Observatoire astronomique de Strasbourg, Universit\'e de Strasbourg, CNRS, UMR 7550, 11 rue de l'Universit\'e, F-67000 Strasbourg, France}
\altaffiltext{2}{Max-Planck-Institut f\"ur Astronomie, K\"onigstuhl 17, D-69117 Heidelberg, Germany}
\altaffiltext{3}{Institute for Astronomy, University of Edinburgh, Royal Observatory, Blackford Hill, Edinburgh EH9 3HJ, UK}
\altaffiltext{4}{Department of Astronomy, University of Michigan, 500 Church St., Ann Arbor, MI 48109, USA}
\altaffiltext{5}{Institute for Astronomy, University of Hawaii at Manoa, Honolulu, HI 96822, USA}
\altaffiltext{6}{Department of Physics, Durham University, South Road, Durham DH1 3LE, UK}
\altaffiltext{7}{Department of Astrophysical Sciences, Princeton University, Princeton, NJ 08544, USA}

\begin{abstract}
We present a new satellite in the outer halo of the Galaxy, the first Milky Way satellite found in the stacked photometric catalog of the Panoramic Survey Telescope and Rapid Response System 1 (Pan-STARRS1) Survey. From follow-up photometry obtained with WFI on the MPG/ESO 2.2m telescope, we argue that the object, located at a heliocentric distance of $145\pm17\kpc$, is the most distant Milky Way globular cluster yet known. With a total magnitude of $M_V=-4.3\pm0.2$ and a half-light radius of $20\pm2\pc$, it shares the properties of extended globular clusters found in the outer halo of our Galaxy and the  Andromeda galaxy. The discovery of this distant cluster shows that the full spatial extent of the Milky Way globular cluster system has not yet been fully explored.
\end{abstract}

\keywords{Local Group --- globular cluster: individual: PSO J174.0675-10.8774}

\section{Introduction}
As compact stellar systems that can be discovered at large distances, globular clusters (GCs) located in the outskirts of massive galaxies are valuable tracers of the hierarchical formation of their host \citep{searle78,law10,mackey10b,pota13}. In particular, the detailed study of nearby GCs has shown that the most distant Milky Way (MW) GCs preferentially belong to the class of so-called `young halo' clusters \citep{mackey05}. These clusters are expected to have formed in dwarf galaxies (DGs) that were later accreted onto our Galaxy and destroyed by tidal forces \citep{dotter11}. Young halo GCs are preferentially younger (8--12\Gyr\ old), more metal-rich ($\FeH\sim-1.5$), and more extended than other halo GCs \citep{dotter10}. Similar conclusions are reached for some GCs in the outskirts of M31 \citep{mackey13}.

Although large sky surveys like the Sloan Digital Sky Survey (SDSS) have transformed our view of the MW satellite DG system \citep[e.g.][]{willman05a,belokurov07a}, only a handful of GCs were discovered within the survey \citep{koposov07,belokurov10,balbinot13}; all of these are extremely faint ($M_V\simgt-2.0$) and within the inner $\sim$ 60 kpc of the halo. Additionally two other GCs within this spatial regime were found on sky survey plates (Pyxis and Whiting 1;  \citealt{Irwin1995}, \citealt{whiting02}), yet no new distant MW GC has been discovered since the searches that led to the discovery of the Palomar clusters \citep[e.g][]{arp60} and AM-1, the most distant MW GC to date at a galactocentric distance of $\sim125 \kpc$ \citep{madore79,aaronson84,Dotter08}.

Here, we report the discovery of the most distant MW GC, PSO J174.0675-10.8774, found in the Pan-STARRS~1 (PS1) $3\pi$ photometric survey. We argue that PSO J174.0675-10.8774 shares the properties of known young halo GCs. The letter is structured as follows: in Section~2 we briefly describe the PS1 survey and the satellite search that led to the discovery of PSO J174.0675-10.8774. Section~3 focusses on the analysis of follow-up wide-field imager data. We derive the properties of the cluster in Section~4 and discuss their implication on the nature of the cluster in Section~5.

Whenever necessary, magnitudes are dereddened using the \citet{schlegel98} maps, assuming the extinction coefficients of \citet{schlafly11}. We also assume a heliocentric distance of $8\kpc$ to the Galactic center.

We note that the same stellar system was discovered independently by \citet{belokurov14b} using VST ATLAS data. In their pre-print, these authors favor a DG classification for this object.

\section{The $3\pi$ PS1 Survey and discovery}
The PS1 3$\pi$ Survey (K. Chambers et al., in preparation) targets three quarters of the sky ($\delta>-30\deg$) in five photometric bands, $g_\mathrm{P1}$, $r_\mathrm{P1}$, $i_\mathrm{P1}$, $z_\mathrm{P1}$, $y_\mathrm{P1}$, with the 1.8m PS1 telescope, located in Haleakala, Hawaii \citep{tonry12}. The sky is surveyed with a 1.4-gigapixel camera covering a 3.3-degree field of view, which, combined with short exposures four times per year per filter over the course of 3.5 years, yields a deep, panoptic view of the MW's surroundings. Once the individual frames are downloaded from the summit, they are automatically processed with the Image Processing Pipeline \citep{magnier06,magnier07,magnier08} to generate a photometric catalogue. The PS1 survey is currently at the stage where stacked photometry and images have become available, reaching a similar depth to the SDSS in $g_\mathrm{P1}$ and reaching deeper magnitude for $r_\mathrm{P1}$ ($\sim0.5$~magnitude) and $i_\mathrm{P1}$ ($\sim1$~magnitude; \citealt{metcalfe13}). The PS1 survey represents a significant increase of the search volume for small scale stellar systems that orbit the MW.

PSO J174.0675-10.8774 was discovered in an on-going search for small-scale substructures in the PS1 data. A comprehensive paper about this search is in preparation (B. Laevens et al., in preparation), but we provide here a broad outline of the discovery method. Inspired by previous searches for stellar overdensities \citep{koposov07,walsh09}, we adopt a hybrid method of the two aforementioned papers. First, we identify stars which have colors and magnitudes compatible with old, metal-poor stars possibly pertaining to a GC or DG by constructing color-magnitude masks from typical isochrones describing such systems and accounting for the possible distance to the system. The subsample of stars for a given distance is then convolved with Gaussian kernels. A positive convolution is performed on the data with a Gaussian of $4'$ or $8'$ dispersion, tailored to the size of typical MW satellites, while a convolution with a large negative Gaussian kernel accounts for the slowly-varying contamination \citep{koposov09}. After this two-step process, the resultant differential density maps track localized stellar over- and underdensities over the PS1 footprint, which we then translate into detection significance by comparison with the local density values. This led to the discovery of PSO J174.0675-10.8774 as an unambiguous $10\sigma$ detection.

\begin{figure*}
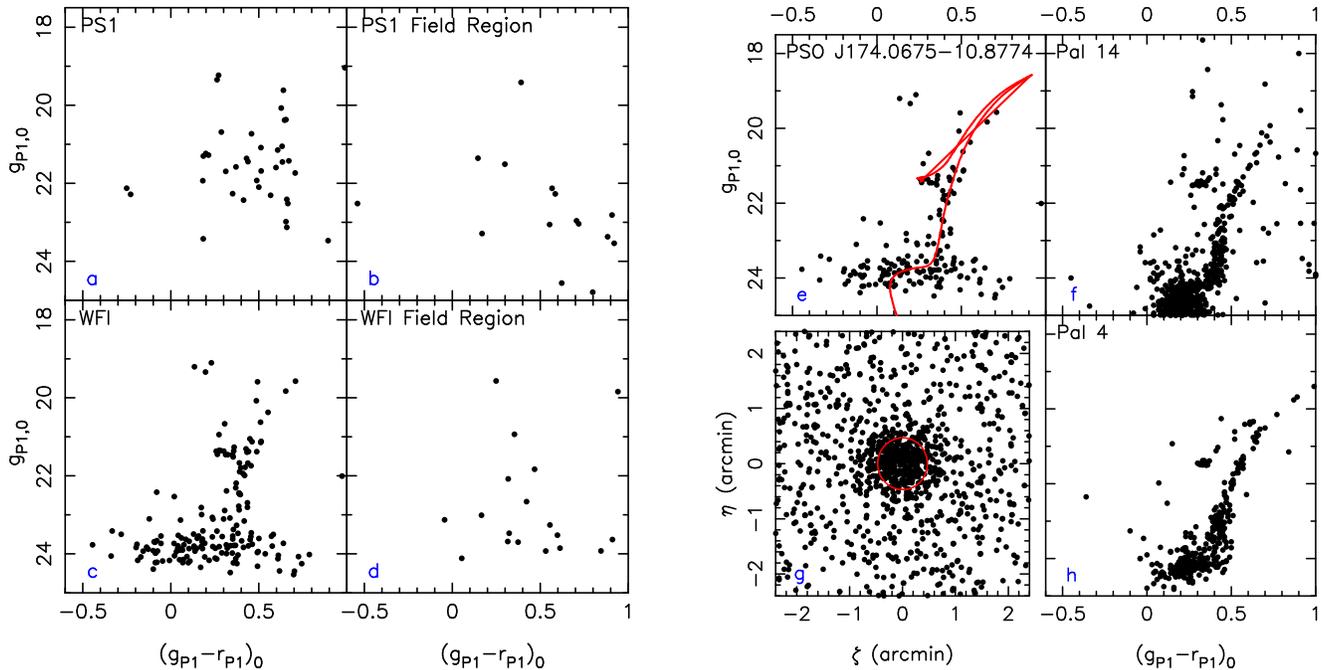

\begin{center}
\includegraphics[width=0.49\hsize,angle=270]{fig1a.ps}\hspace{1cm}
\includegraphics[width=0.49\hsize,angle=270]{fig1b.ps}
\caption{\label{CMDs}\emph{Left panels:} \textbf{a)}: The CMD of PS1 stars within 3 half-light radii of the centroid of PSO J174.0675-10.8774. \textbf{b)}: The CMD of stars of a nearby field region (8.3 arcminutes North-East away) of the same coverage. \textbf{c)}: WFI stars within 2 half-light radii of the centroid of PSO J174.0675-10.8774. \textbf{d)}: WFI stars within 2 half-light radii of the aforementioned field region. The CMD of the new GC shows a clear RGB, HB, and grazes its main sequence turn-off at the faint end. \emph{Right panels:} \textbf{e)}: The CMD of PSO J174.0675-10.8774 with the \textsc{Parsec} isochrone of age $8\Gyr$ and $\FeH=-1.9$ that matches the shape and location of the RGB, HB, and main sequence turn off. \textbf{g)}: The spatial distribution of the WFI stars, displaying the unambiguous overdensity produced by PSO J174.0675-10.8774. The red circle shows the region within 2 half-light radii of its centroid. \textbf{f, h)}: CMDs within two half-light radii of young outer halo GCs Pal 14 and Pal 4, shifted to the distance of the new GC. This photometry is taken from \citet{saha05}. The CMDs of these two stellar systems show many similarities with that of PSO J174.0675-10.8774, especially their red HBs and sparsely populated RGBs.}
\end{center}
\end{figure*}

At this location, the PS1 stacked images reveal the presence of a compact stellar system, which is emphasized by the distribution of PS1 sources on the sky. In panel a of Figure~\ref{CMDs}, the CMD of this compact overdensity confirms that these stars are blue ($(g_\mathrm{P1}-r_\mathrm{P1})_0\simlt0.8$) and more numerous than in a field region (panel b) of the same coverage. Despite the detection of the object in the PS1 data, these are too shallow to reliably determine the properties of the system, which motivated us to gather deeper photometric data.

\section{Follow-Up}
\begin{figure}
\begin{center}
\includegraphics[width=\hsize,angle=0]{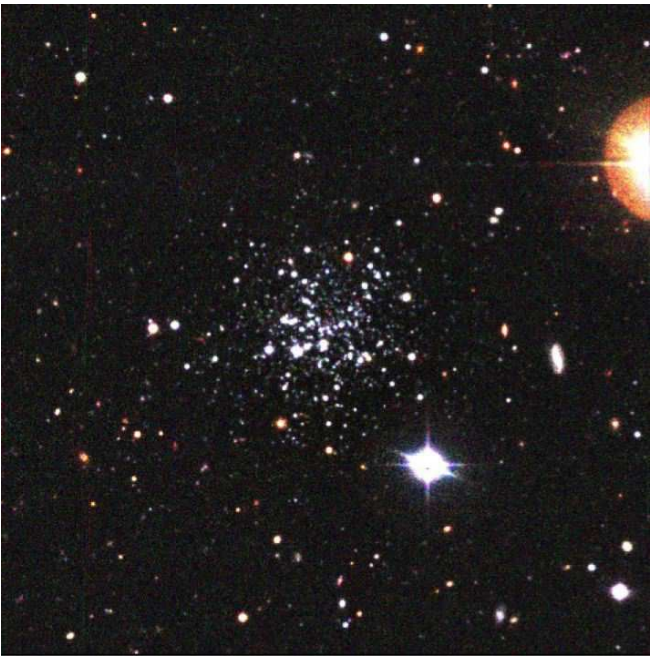}
\caption{\label{BVRimage} $BVR$ image of PSO J174.0675-10.8774 built from the stacked WFI images. The image is $4'\times4'$, north is to the top and east to the left.}
\end{center}
\end{figure}

During the night of December 31, 2013--January 1st, 2014, we obtained follow-up imaging with the Wide Field Imager (WFI) on the MPG/ESO 2.2m telescope located in La Silla, Chile. Equipped with 8 CCDs the camera has a field of view of $34'\times33'$. Given the small angular size of PSO J174.0675-10.8774, we focus here on the photometry from CCD~1, which contains the full extent of the new stellar system. We obtained imaging in the $B$ (five six-minute dithered sub-exposures), $V$, and $R$ bands (three seven-minute dithered sub-exposures) in good seeing conditions with a median image quality of $0.9"$. The individual sub-exposures were bias-subtracted, flat-fielded, and cleared of cosmic rays. The initial astrometric solution for sub-exposures was obtained using the Astrometry.net package \citep{lang10}, with the PS1 catalog acting as the astrometric reference catalog. We then used the software packages SExtractor and SCAMP \citep{bertin06} to align each individual frame to the frame with the best seeing before constructing composite images for each band by average combining them with SWARP \citep{bertin06}. The resulting $BVR$ image of PSO J174.0675-10.8774 is shown in Figure~\ref{BVRimage} and unequivocally confirms that it is a compact stellar system.

We performed the photometry using the {\sc daophot/allstar/allframe} suite of programs \citep{stetson94} as described in \citet[][E.J. Bernard, in preparation]{martin13}. A point-spread-function model was built for each individual frame and the flux was extracted by applying the model to fixed positions of stars as measured from the deep $BVR$ image. The instrumental magnitudes were then calibrated to the first sub-exposure in a band, averaged out. We use the exiquisite photometry of the single epoch PS1 data \citep{schlafly12} to derive our own transformations from the WFI filters to the PS1 $g_\mathrm{P1}$- and $r_\mathrm{P1}$-band magnitudes.

The final WFI photometry reaches $\sim1.5$ magnitudes deeper than the stacked PS1 data, as shown in panel c of Figure~\ref{CMDs}. The CMD within two half-light radii of PSO J174.0675-10.8774 (see section 4 for the determination of the structural parameters) unveils a sparse red giant branch (RGB), a red horizontal branch (HB), and the beginning of the system's main sequence turn off at the magnitude limit of the data. All of these features are absent from a field region (panel d) of similar coverage shown for comparison. The spatial distribution of stars in the WFI data is shown in panel g.

\section{Properties of the stellar system}
A search for possible RR Lyrae stars varying in the multi-epoch PS1 photometry revealed no good candidate. However, although the CMD of PSO J174.0675-10.8774 is only sparsely populated, the presence of a well defined (red) HB allows for an accurate determination of the heliocentric distance to the system. We therefore measure the median apparent magnitude of the HB through a Monte Carlo resampling of the magnitude of stars in this part of the CMD, yielding $m_{g\mathrm{P1}} = 21.41\pm 0.06$. To determine the absolute magnitude of the red horizontal branch in this band, we extract from the PS1 stacked photometry database the photometry of four GCs that share the morphological properties and CMD of PSO J174.0675-10.8774 (Pal~3, Pal~4, Pal~5, and Pal~14). We repeat the same procedure on the red HB of these systems, shifted by their distance modulus as listed in \citet{harris10}. The weighted average of these four HB's absolute magnitude measurements yields $M_{g\mathrm{P1}}=0.60\pm0.10$. The distance modulus of PSO J174.0675-10.8774 is therefore $(m-M)_0=20.81\pm 0.12$, yielding both a heliocentric and a Galactocentric distance of $145\pm17 \kpc$.

Having pinned down the distance to the system, we can now estimate its age and metallicity through a comparison with a set of isochrones. Since the photometric data does not reach the main sequence of the cluster, these estimates should be taken with caution, but they nevertheless give a broad understanding of the age and metallicity of the cluster. Fixing the distance modulus at $(m-M)_0=20.81$, we cycle through the \textsc{Parsec} isochrones for a metallicity range $-2.3<\FeH<-1.5$ and an age between 8 and $13\Gyr$ \citep{bressan12}. The top-right panel (e) of Figure~\ref{CMDs} presents the best fit to the data, obtained for the isochrone with an age of $8 \Gyr$ and $\FeH=-1.9$. Allowing the distance to vary within the formal distance uncertainties, we obtain a metallicity and age range of 8--10 Gyrs and $\FeH$ varies between $-1.5$ and $-1.9$. These are fairly typical properties of young, outer halo GCs and this impression is further bolstered by panels f and h of Figure~\ref{CMDs}, which present literature photometry \citep{saha05} of Pal~14 (11.3\Gyr, $\FeH=-1.5$; \citealt{dotter11}) and Pal~4 (10.9\Gyr, $\FeH=-1.3$) shifted to the distance of the new GC. The CMDs of the three GCs exhibit similar features, with sparsely populated RGBs and red HBs.

\begin{figure}
\begin{center}
\includegraphics[width=0.72\hsize,angle=270]{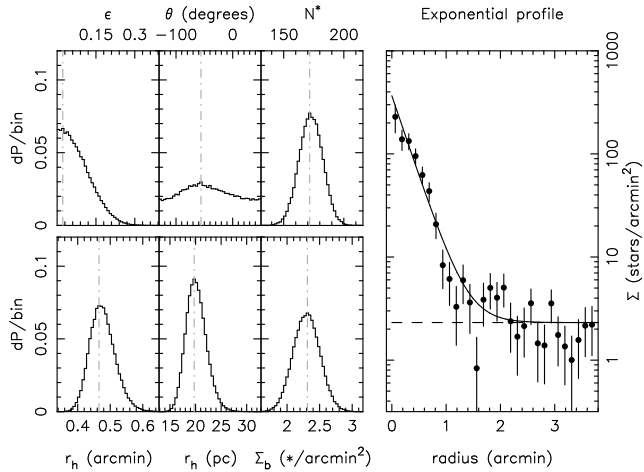}
\caption{\label{profile}The left-hand panels show the pdf of the structural parameters for the exponential radial density model. From top-left to bottom-right, the parameters correspond to the ellipticity ($\epsilon$), the position angle ($\theta$), the number of stars in the GC for our CMD selection ($N^*$), the angular and physical half-light radii ($r_h$), and the field density ($\Sigma_\mathrm{b}$). The right-most panel compares the favored radial distribution profile (full line) and the data, binned following with the favored structural parameters (dots). The field density is shown as the dashed line and the uncertainties on the data point were calculated assuming Poisson uncertainties.}
\end{center}
\end{figure}

\begin{table}
\caption{\label{properties}Properties of PSO J174.0675-10.8774}
\begin{tabular}{cc}
\hline
$\alpha$ (J2000) & 11:36:16.2\\
$\delta$ (J2000) & $-$10:52:38.8\\
$\ell$ & $274.8\deg$\\
$b$ & $+47.8\deg$\\
Distance Modulus & $21.81\pm0.12$\\
Heliocentric Distance & $145\pm17\kpc$\\
Galactocentric Distance & $145 \pm17\kpc$ \\
$M_{V}$ & $-4.3\pm0.2$\\
\FeH & $\sim-1.9$\\
Age & $\sim8\Gyr$\\
$E(B-V)^{a}$ & 0.026\\
\hline
 & Exponential profile\\
Ellipticity & $0.00^{+0.10}_{-0.00}$\\
Position angle (from N to E) & $-56^{+88}_{-68}$$\deg$\\
$r_{h}$ & $0.47^{+0.04}_{-0.03}$$'$\\
& $20\pm2\pc$\\
\hline
 & Plummer profile\\
Ellipticity & $0.0^{+0.11}_{-0.00}$\\
Position angle (from N to E) & $-64\pm64\deg$\\
$r_{h}$ & $0.52\pm0.04'$\\
& $22\pm2\pc$\\
\hline
 & King profile\\

Ellipticity & $0.00^{+0.11}_{-0.00}$\\
Position angle (from N to E) & $+70^{+48}_{-36}$$\deg$\\
$r_{c}$ & $0.61\pm0.18'$\\
& $24^{+9}_{-6}$$\pc$\\
$r_{t}$ & $1.30^{+0.19}_{-0.12}$$'$\\
& $56^{+9}_{-6}$$\pc$\\

\hline
$^a$ from \citet{schlegel98} &\\
\end{tabular}
\end{table}

To determine the structural parameters of PSO J174.0675-10.8774, we use a variant of the technique presented in \citet{martin08b}, updated to allow for a full Markov Chain Monte Carlo treatment. Briefly, the algorithm uses the location of every single star in the WFI data set to calculate the likelihood of a family of radial profiles with flattening and a constant background. The parameters are: the centroid of the system, its ellipticity\footnote{The ellipticity is here defined as $1-b/a$ with $a$ and $b$ the major and minor axis scale lengths, respectively.}, the position angle of its major axis from N to E, the number of stars in the system, and one or two scale parameters. We use three different families of radial density models (exponential, Plummer, and King), for which the scale parameters are the half-light radius, the Plummer radius, and the core and King radii, respectively. The resulting structural parameters are listed in Table~1. Figure~\ref{profile} also shows the probability distribution function (pdf) of the exponential model parameters, as well as the comparison between the favored radial distribution profile and the data, binned with the favored structural parameter model. They show a very good agreement, testament to the quality of the structural parameter's inference. The results for the Plummer profile and the King profile are equally good. From the determination of the structural parameters, we find that PSO J174.0675-10.8774 is a round, compact system with a half-light radius of only $0.47^{+0.04}_{-0.03}$$'$ or $20\pm2\pc$ at the distance of $145\pm17\kpc$.

The absolute magnitude of PSO J174.0675-10.8774 is also derived using the technique presented in \citet{martin08b}, which accounts for `CMD shot-noise' that stems from the impact of the sparsely populated CMD on the derivation of its total magnitude. Using the favored \textsc{Parsec} isochrone presented above (8 \Gyr, $\FeH=-1.9$) and its associated luminosity function shifted to the distance of the cluster, we build a color-magnitude pdf of where PSO J174.0675-10.8774 stars should lie given the photometric uncertainties. We then populate a mock CMD from this pdf until the number of stars brighter than $r_\mathrm{P1}=23.5$ equals the number of stars, $N^*$, that we determine from the estimate of the structural parameters for the same region of the CMD. Summing up the flux of all the stars in the mock CMD yields the absolute magnitude of this realization. Finally, we iterate this procedure 300 times, accounting for the uncertainties on the heliocentric distance and on $N^*$, as well as the shot-noise uncertainties that come from randomly populating the mock CMDs. This procedure yields absolute magnitudes of $M_g=-4.02\pm 0.22$ and $M_r=-4.49\pm 0.25$, which converts to $M_V=-4.3\pm0.2$ when using the \citet{tonry12} color transformations. The total luminosity of the system is $4.5 \pm 0.7 \times 10^3 L_\sun$, which, assuming a typical mass-to-light ratio in solar units of $\sim1.5$ for GCs of this metallicity \citep[e.g.][]{strader11}, yields a total stellar mass of $\sim6.8\pm1.1\times 10^3 M_\sun$.

\section{Discussion and Conclusion}

\begin{figure*}
\begin{center}
\includegraphics[width=0.4\hsize,angle=270]{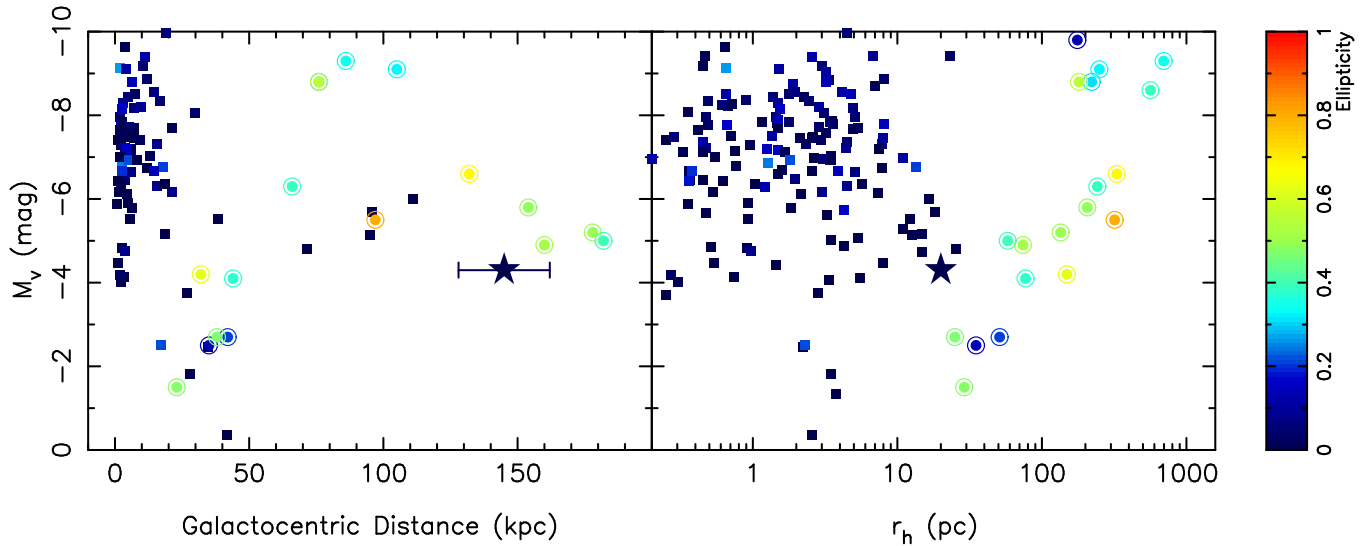}
\caption{\label{Faint_GCs_Props}The distribution of MW satellites in the distance--magnitude space (left) and the size--magnitude space (right). GCs are shown as squares, DGs are shown as circles, and PSO J174.0675-10.8774 is represented by the large star symbol. The color scale indicates the ellipticity of the various satellites. PSO J174.0675-10.8774Õs ellipticity and half-light radius show very similar values to those of other MW GCs. The data on the GCs were taken from \citet{harris10} and these on the DGs from \citet{mcconnachie12}.}
\end{center}
\end{figure*}

We have presented a new stellar system found within the PS1 $3\pi$ data in the outer halo of the MW. Located at a heliocentric distance of $145\pm17\kpc$, this system is rather faint ($M_V=-4.3\pm0.2$), compact ($r_h=20\pm2\pc$), round ($\epsilon=0.0^{+0.10}_{-0.00}$), younger than the oldest GCs ($\sim8\Gyr$), and fairly metal-poor ($\FeH\sim-1.9$). Figure~\ref{Faint_GCs_Props} compares the properties of PSO J174.0675-10.8774 with MW GCs and DGs. Although it is slightly more distant than any known MW GC, all the other properties of this system are similar to those of outer halo GC systems \citep[e.g.][]{mackey04,mackey05}. In particular, the size and the roundness of the system differentiate it from DGs at the same magnitude, which are all larger than at least $r_h=60\pc$ and tend to favor elliptical radial density profiles \citep{martin08b,sand12}. Furthermore, the age and metallicity we determine from the comparison with isochrones are quite typical of young, outer halo GCs, as made evident by the direct comparison of PSO J174.0675-10.8774's CMD with that of Pal~4, and Pal~14 (Figure~\ref{CMDs}e, f, and h).

Beyond our local environment, this new stellar system is also similar to the population of extended GCs recently found in the vicinity of M31, many of which lie on faint stellar streams \citep{huxor05,mackey10b}. Those systems are similarly round, rather fluffy for GCs, but compact for DGs, and can also exhibit red horizontal branches \citep[e.g][]{mackey06,huxor11,mackey13}. The discovery of this most distant MW GC supports the idea that very remote GCs are a common feature of large spiral galaxies as has been shown in M31 \citep{huxor11}, in M33 \citep{cockcroft11} and M81 \citep{jang12}. Further investigation in the vicinity of this object would therefore be interesting in the context of stellar streams and their association with GCs.

After the work for this letter had been completed, we learned of the independent discovery of this object by \citet{belokurov14b} from the VST ATLAS survey and follow-up data. Their determination of the system's properties are consistent with ours, even though they derive a slightly larger distance ($\sim170\kpc$) and size ($r_h\sim0.6'=30\pc$ at their distance). Their interpretation of the nature of the system nevertheless differs from ours as they favor a scenario in which the system is a DG. Their conclusion is partly driven by the larger size they measure, but also by their interpretation of a handful of blue stars being blue loop stars indicative of recent star formation. We do indeed retrieve these stars in our data set; however, we are cautious as to their interpretation. Performing our structural parameter analysis on these blue stars only yields a detection of an overdensity at the $\sim$ 2--$3\sigma$ detection level. Furthermore, if these were truly blue loop stars, one would expect the presence of a higher number of their low-mass analogs at fainter magnitudes and bluer colors, which do not appear in the CMD. Finally, it cannot be ruled out that the two blue stars residing barely above the red HB could in fact be AGB stars. Taking into account the derived properties of this stellar system, we rather conclude that we are in the presence of a distant GC and not of a faint and extremely compact DG. We stress that this conclusion must be confirmed or infirmed with radial velocities through follow-up spectroscopy of these blue loop stars, thus ascertaining the definite nature of this new distant satellite.

\acknowledgments

We would like to thank the editor and referee for their celerity. We wish to thank P. Bianchini and M. A. Norris for fruitful discussions about globular clusters and R. A. Ibata for a careful reading of the manuscript. B.P.M.L. acknowledges funding through a 2012 Strasbourg IDEX (Initiative d'Excellence) grant, awarded by the French ministry of education. N.F.M. and B.P.M.L. gratefully acknowledges the CNRS for support through PICS project PICS06183. N.F.M., H.-W.R. \& E.F.S. acknowledge support by the DFG through the SFB 881 (A3). E.F.B. and C.T.S. acknowledge support from NSF grant AST 1008342.

The Pan-STARRS1 Surveys (PS1) have been made possible through contributions of the Institute for Astronomy, the University of Hawaii, the Pan-STARRS Project Office, the Max-Planck Society and its participating institutes, the Max Planck Institute for Astronomy, Heidelberg and the Max Planck Institute for Extraterrestrial Physics, Garching, the Johns Hopkins University, Durham University, the University of Edinburgh, Queen's University Belfast, the Harvard-Smithsonian Center for Astrophysics, the Las Cumbres Observatory Global Telescope Network Incorporated, the National Central University of Taiwan, the Space Telescope Science Institute, the National Aeronautics and Space Administration under Grant No. NNX08AR22G issued through the Planetary Science Division of the NASA Science Mission Directorate, the National Science Foundation under Grant No. AST-1238877, the University of Maryland, and Eotvos Lorand University (ELTE).



\end{document}